\documentclass[a4paper]{article}
\usepackage{RR}
\usepackage{hyperref}
\RRdate{June 2006} 
\RRauthor{Bin Bin Chen \and Pascale Primet} 
\authorhead{B. Chen \& P. Primet}
\RRtitle{Un cadre flexible de r\'eservation de bande passante pour
les transferts massifs dans les r\'eseaux de grille}
\RRetitle{A Flexible Bandwidth Reservation Framework for Bulk Data
Transfers in Grid Networks} 

\titlehead{Flexible bandwidth reservation for bulk transfer}
\RRresume{Dans les r\'eseaux de grilles, les ressources
distribu\'ees sont interconnect\'ees par des r\'eseaux longues
distance pour ex\'ecuter des applications intensives de calcul ou
de traitement de donn\'ees, qui n\'ecessitent des transferts
fiables et efficaces de volumes de donn\'ees de l'ordre de
plusieurs gigaoctets ou teroctets. Le transferts massifs dans les
grilles, contrairement au trafic ``best effort'' de l'Internet,
requierent un service de r\'eservation de bande-passante. Les
sch\'emas de r\'eservation existants, tels RSVP, ont \'et\'e con\c
cus pour du trafic temps-r\'eel et pour lequel on sp\'ecifie un
d\'ebit r\'eserv\'e, une date de d\'ebut de transfert mais on ne
pr\'ecise pas la dur\'ee. En comparaison, les transferts massifs
de grilles sont d\'efinis en termes de volumes et de date limite,
ce qui offre plus d'informations et autorise des sch\'emas de
r\'eservation plus flexibles. Le d\'ebut effectif du transfert
peut \^etre choisi, une r\'eservation pour une m\^eme requ\^ete
peut \^etre divis\'es en plusieurs intervals avec des d\'ebits
r\'eserv\'es diff\'erents. Nous d\'efinissons un cadre flexible de
r\'eservation de bande passante \`a l'aide d'une alg\`ebre de
fonctions temps-d\'ebit et identifions une s\'erie de familles de
sch\'emas de r\'eservation, que nous nommons FixTime-FixRate,
FixTime-FlexRate, FlexTime FlexRate, et Multi-Interval,
pr\'esentant une g\'en\'eralit\'e et un potentiel de performance
croissants. Des heuristiques simples sont utilis\'ees pour
s\'electionner un sch\'ema repr\'esentatif dans chaque famille
pour comparer les performances. Les r\'esultats de simulation
montrent que l'augmentation de la flexibilit\'e peut
potentiellement augmenter les performances du syst\`eme, minimiser
la probabilit\'e de blocage et la dur\'ee moyenne des flux. Nous
discutons aussi de l'implantation distribu\'ee du cadre
propos\'e.}

\RRabstract{In grid networks, distributed resources are
interconnected by wide area network to support compute and
data-intensive applications, which require reliable and efficient
transfer of gigabits (even terabits) of data. Different from
best-effort traffic in Internet, bulk data transfer in grid
requires bandwidth reservation as a fundamental service. Existing
reservation schemes such as RSVP are designed for real-time
traffic specified by reservation rate, transfer start time but
with unknown lifetime. In comparison, bulk data transfer requests
are defined in terms of volume and deadline, which provide more
information, and allow more flexibility in reservation schemes,
i.e., transfer start time can be flexibly chosen, and reservation
for a single request can be divided into multiple intervals with
different reservation rates. We define a flexible reservation
framework using time-rate function algebra, and identify a series
of practical reservation scheme families with increasing
generality and potential performance, namely, FixTime-FixRate,
FixTime-FlexRate, FlexTime-FlexRate, and Multi-Interval. Simple
heuristics are used to select representative scheme from each
family for performance comparison. Simulation results show that
the increasing flexibility can potentially improve system
performance, minimizing both blocking probability and mean flow
time. We also discuss the distributed implementation of proposed
framework.} 

\RRmotcle{R\'eservation, grille, transferts massifs,
flexibilit\'e}

\RRkeyword{Reservation, grid, bulk data transfer, flexibility} 
\RRprojets{RESO} 
\RRtheme{\THNum} 
\URSophia 
\begin{document}
\makeRR   

\section{Introduction}

Grid computing is a promising technology that brings together
large collection of geographically distributed resources (e.g.,
computing, storage, visualization, etc.) to build a very high
performance computing environment for compute and data-intensive
applications \cite{foster04}. Grid networks connect multiple
sites, each comprising a number of processors, storage systems,
databases, scientific instruments, and etc. In grid applications,
like experimental analysis and simulations in high-energy physics,
climate modeling, earthquake engineering, drug design, and
astronomy, massive datasets must be shared by a community of
researchers distributed in different sites. These researchers
transfer large subsets of data across network for processing. The
\emph{volume} of dataset can usually be determined from task
specification, and a strict \emph{deadline} is often specified to
guarantee in-time completion of the whole task, also to enforce
efficient use of expensive grid resources, not only network
bandwidth, but also the co-allocated CPUs, disks, and etc.

While Internet bulk data transfer works well with best-effort
service, high-performance grid applications require bandwidth
reservation for bulk data transfer as a fundamental service.
Besides strict deadline requirement and expensive co-allocated
resources as we discussed above, the \emph{smaller multiplexing
level} of grid networks compared to Internet also serves as a main
driving force for bandwidth reservation. In Internet, the source
access rates are generally much smaller ($2Mbps$ for DSL lines)
than the backbone link capacity (hundreds to thousands of Mbps,
say). Coexistence of many active flows in a single link smoothes
the variation of arrival demands due to the law of large number,
and the link is not a bottleneck until demand attains above $90\%$
of its capacity \cite{roberts04survey}. Thus no proactive
admission control is used in Internet for bulk data transfer.
Instead, distributed transport protocols, such as TCP, are used to
statistically share available bandwidth among flows in a ``fair''
way. Contrarily, in grid context, the capacity of a single source
($c=1Gbps$) is comparable to the capacity of bottleneck link. For
a system with small multiplexing level, if no pro-active admission
control is applied, burst of load greatly deteriorates the system
performance.

A concrete example is given in Section \ref{sec:motivation} to
demonstrate the importance of resource reservation for grid
networks. Through the example, we also show that existing
RSVP-type framework is not flexible enough for bulk data transfer
reservation. In Section \ref{sec:formalization}, we define a
flexible reservation framework using time-rate function algebra.
Section \ref{sec:schemes} identifies a series of practical
reservation scheme families with increasing generality, and we use
simple heuristics to select representative scheme from each
family. In Section \ref{sec:eval}, simulation result of chosen
schemes are presented and the impact of flexibility is analyzed. A
distributed architecture is proposed in Section
\ref{sec:architecture}. In Section \ref{sec:relate}, we briefly
review related works on bandwidth reservation. Finally, we
conclude in Section \ref{sec:conclusion}.
\newpage

\section{Motivation}
\label{sec:motivation}

\begin{figure}[!ht]
\centering
\includegraphics[height=60mm]{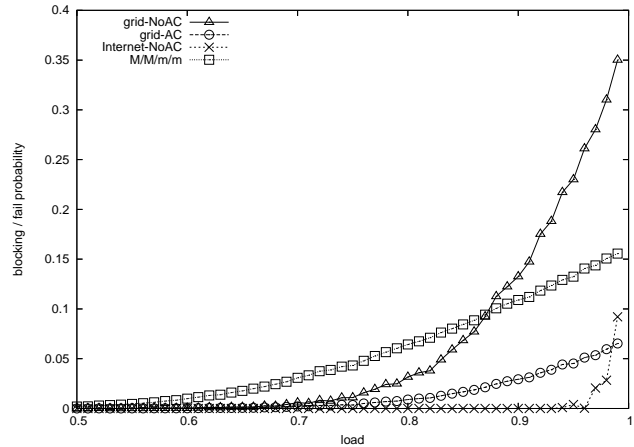}
\caption{Fail/block probability under different multiplexing
level} \label{img:multiplexing}
\end{figure}

In Figure \ref{img:multiplexing}, we simulate a single link with
capacity $C$. Bulk data transfer requests arrive according to a
Poisson process with parameter $\lambda$. Request volume is
independent of arrival time, and follows an exponential
distribution with parameter $\mu$. Simulations with other arrival
processes and traffic volume distributiones reveal similar trend,
which are not presented here for brevity. Load $\rho=\lambda / (C
* \mu)$. Requests have maximal transfer rate $R_{max}$. In
Internet setting $R_{max}^{Internet}=C/100$, and in grid setting
$R_{max}^{grid}=C/10$. \emph{Ideal transport protocol} is assumed,
so that if there are no more than $C/R_{max}$ active flows, all of
them transfer at full rate $R_{max}$. If there are $n > C/R_{max}$
active flows, they all transfer at rate $C/n$. A request with
volume $v$ ``fails'' and immediately terminates, if it does not
complete transfer within $v/R_{min}$ time, where $R_{min}\leq
R_{max}$ is the expected average throughput of the request (in
this example $R_{min}=R_{max}/2$ for all requests). In
\emph{Internet-NoAC} setting (AC stands for ``Admission
Control''), the fail probability is low until load $\rho$ attains
above $95\%$. In \emph{grid-NoAC} setting, however, the fail
probability is nonneglectable even under a medium load, and it
deteriorates rapidly as load increases. Thus we consider using a
simple reservation scheme, which enforces requests to reserve
$R_{min}$ bandwidth when they arrive, so that all accepted
requests are guaranteed to complete before deadline (fail
probability is $0$). Requests are blocked if the number of active
reservations reaches $C/R_{exp}$. This kind of reservation can be
supported by existing reservation schemes, for example, RSVP
\cite{braden97}. In \emph{grid-AC} setting, we still assume
\emph{ideal transport protocol}, i.e., accepted requests are able
to fairly share unreserved capacity in addition to their reserved
bandwidth. Block probability of \emph{grid-AC} setting is much
lower than fail probability of \emph{grid-NoAC} setting.

In Figure \ref{img:multiplexing}, we also plot a variation of
\emph{grid-AC} setting in which flows can only use reserved
bandwidth. With this \emph{dull transport protocol} assumption,
the link can be modeled as a standard $M/M/m/m$ queuing system
with $m=C/R_{min}=10$. Comparing this $M/M/m/m$ setting against
\emph{grid-NoAC} setting, simple reservation scheme with dull
transport protocol can still outperform no admission control
setting with ideal transport protocol when load is relatively
high. This again demonstrates the benefit of reservation.
Meanwhile, the big performance gap between $M/M/m/m$ setting and
\emph{grid-AC} setting shows that when transport protocol is dull,
a RSVP-type reservation does not fully exploit the system's
capacity. The transport protocol design for high speed network is
still an ongoing research. Complementarily, we consider how to
improve system's performance by using more flexible reservation
schemes in this paper.

RSVP is designed for real-time traffic which normally requests for
a specified value of bandwidth from a fixed start time. Their
lifetime is unknown, thus reservation remains in effect for an
indefinite duration until explicit ``Teardown'' signal is issued
or soft state expires. In stead, bulk data transfer requests are
specified by \emph{volume} and \emph{deadline}. This allows more
flexibility in the design of reservation schemes. As volume is
known, the completion time can be calculated by scheduler and kept
in time-indexed reservation states. If there is not enough
bandwidth at the moment a request arrives, transfer can be
scheduled to start at some future time point as long as it can
complete before deadline. Bandwidth reservation can also comprise
sub-intervals with different reserved rates.

\begin{figure}[!ht]
\centering
\includegraphics[height=50mm]{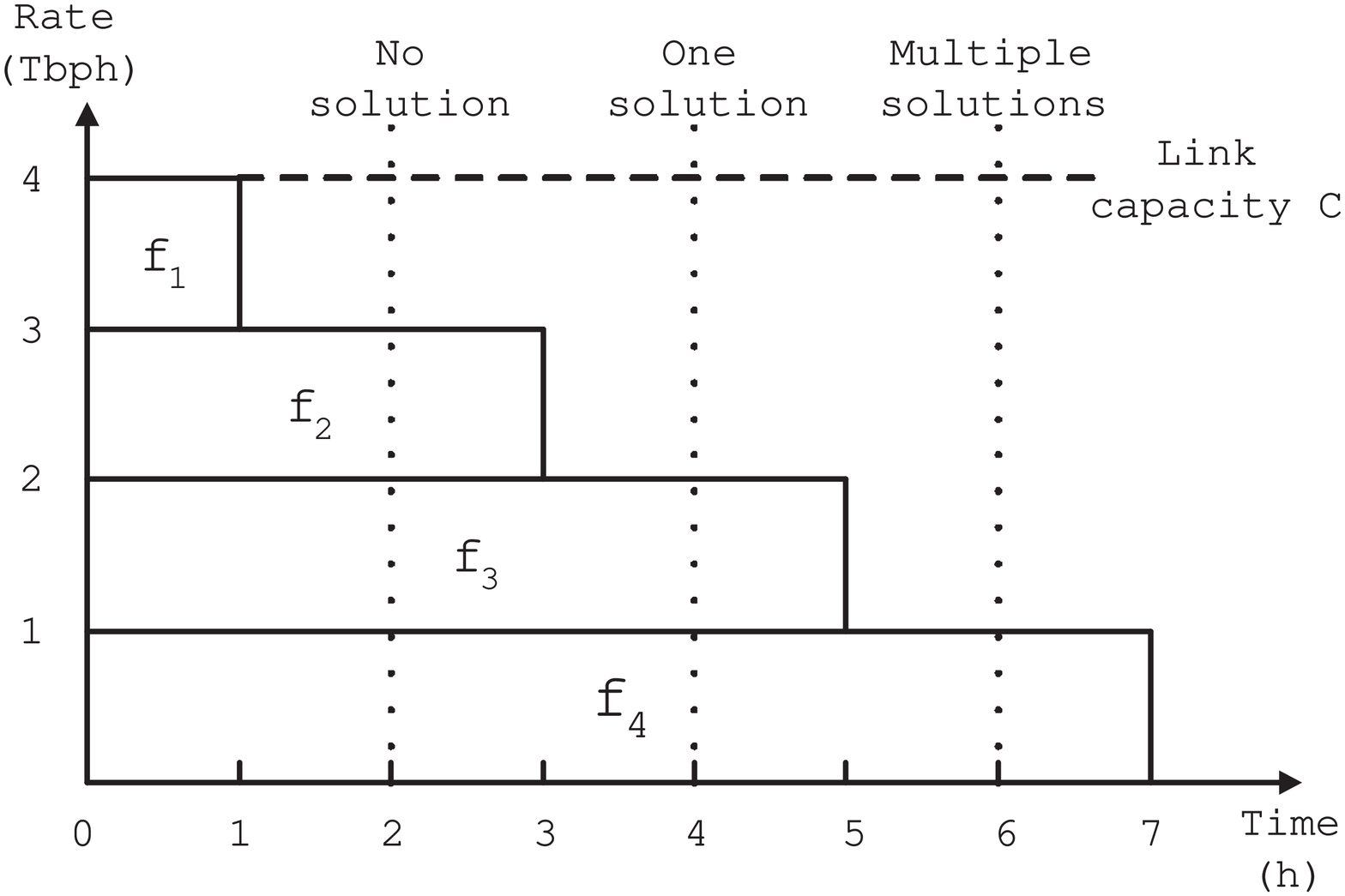}
\caption{Flexible reservation schemes example} \label{img:example}
\end{figure}

Limitation of RSVP-type reservation for bulk data transfer is
illustrated in Figure \ref{img:example}. In this example, we
consider a link with capacity $C=4Tbph$. Requests arrive online
with varying volume, their maximal transfer rate is
$R_{max}=2Tbph$ and their minimum average transfer rate is
$R_{min}=1Tbph$. A request arrives at time $t$ with volume $v$ has
a deadline $t+v/R_{min}$. Assume at current time $0h$, there are
four active reservations each reserving $1Tbph$ bandwidth. Their
termination times are known and marked in the figure. A new
request arrives at $0h$ with volume $v=4Tb$, and its deadline is
$0h + 4Tb/R_{min} = 4h$. Since there is no bandwidth left at time
$0h$, this request will be rejected by RSVP-type reservation
scheme. This unnecessary rejection can be avoided, if we use more
flexible reservation scheme and exploit the time-indexed
reservation state information. A feasible reservation solution is
to reserving $1Tbph$ for the request from time $1h$ (other than
from $0h$) until $3h$, followed by a different reservation rate of
$2Tbph$ until $4h$.

In the case of $v=4Tb$, this is the only solution to accept the
request and guarantee its successful completion without preempting
any existing reservations. However, if the request has volume
$v=2Tb$ and thus deadline $2h$, no feasible solution exists to
accept the new request unless \emph{preemption} is allowed. The
concept of \emph{preemption} is borrowed from job scheduling
literature, which means the modification (including teardown) of
the reservation state of an already-accepted request by system.
Compared to \emph{non-preemptive} schemes, \emph{preemptive
schedulers} enjoy higher decision flexibility which implies
potential performance gain. But they have some drawbacks
including:
\begin{itemize}
\item Dropping accepted request causes more dissatisfaction than
blocking new one; \item Dynamic change (QoS degration) of
reservation state hurts service predictability, which is important
because bandwidth is co-allocated with other resources.
\end{itemize}
Also, it is challenging to design a distributed preemptive
reservation architecture. In this paper, we focus on a
\emph{non-preemptive reservation framework}.

There may be multiple feasible solutions to accept a request, for
example if the request here is with volume $v=6Tb$ and deadline
$6h$. The algorithm to select a solution out of all feasible
solutions depends on the objective functions of reservation
schemes. Besides increasing accept probability, there are other
important performance criteria. Borrowing concept again from job
scheduling, \emph{flow time} is defined as the time between a
request's arrival and its completion. For bulk data transfers,
especially in grid applications, it is desirable to minimize flow
time. Smaller flow time not only improves users' satisfaction, but
also releases all co-allocated resources earlier back to sharing
pool. Fairness among flows is also an important performance
criteria. For example, bulk data transfer may define fairness over
their average throughput. These criteria may be conflicting with
each other. For example, the solution to minimize flow time here
is to reserve $1Tbph$ from $1h$ to $3h$, and $2Tbph$ from $3h$ to
$5h$ so that the request can be finished at $5h$. While the
solution to minimize peak reservation rate is to reserve $1Tbph$
from $1h$ to $3h$, and $4/3Tbph$ from $3h$ to $6h$. Yet another
reasonable solution is to reserve $0.5Tbph$ from $1h$ to $3h$,
$1.5Tbph$ from $3h$ to $5h$, followed by $2Tbph$ ($R_{max}$) from
$5h$ to $6h$, so that the remained bandwidth variation along time
axis is minimized.

It is very difficult (if not totally impossible) to identify the
optimal solution in both off-line and on-line setting. Sometimes
it is preferable to reject a request even when feasible solution
exists. In this paper, we don't emphasis the choice of objective
functions and optimal solutions. Instead, we focus on formalizing
a flexible yet practical solution space, so that a potential
candidate solution will not be missed because of the limitation in
reservation framework flexibility.
\newpage

\section{Flexible reservation framework}
\label{sec:formalization}

\subsection{System model}

We model grid networks as a set of resources interconnected by
wide area network. The underlying communication infrastructure of
grid networks is a complex interconnection of enterprize domains
and public networks that exhibit potential bottlenecks and varying
performance characteristics. For simplicity, we assume a
centralized scheduler manages reservation state vector $L$ for all
links in the system. We will discuss the distributed
implementation in Section \ref{sec:architecture}.

We define a request as a 6-tuple:
\begin{equation}r=(s_r, d_r, v_r, a_r, d_r, R^{max}_r)\end{equation}
As suggested by name, source $s_r$ requests to transfer bulk data
of volume $v_r$ to destination $d_r$. Request arrives at time
$a_r$ and transfer is ready to begin immediately. Transfer should
complete before deadline $d_r$, and $R^{max}_r$ is the maximum
rate that request $r$ can support, constrainted by either link
capacity of end nodes, application or transport protocol.

\begin{figure}[!ht]
\centering
\includegraphics[height=40mm]{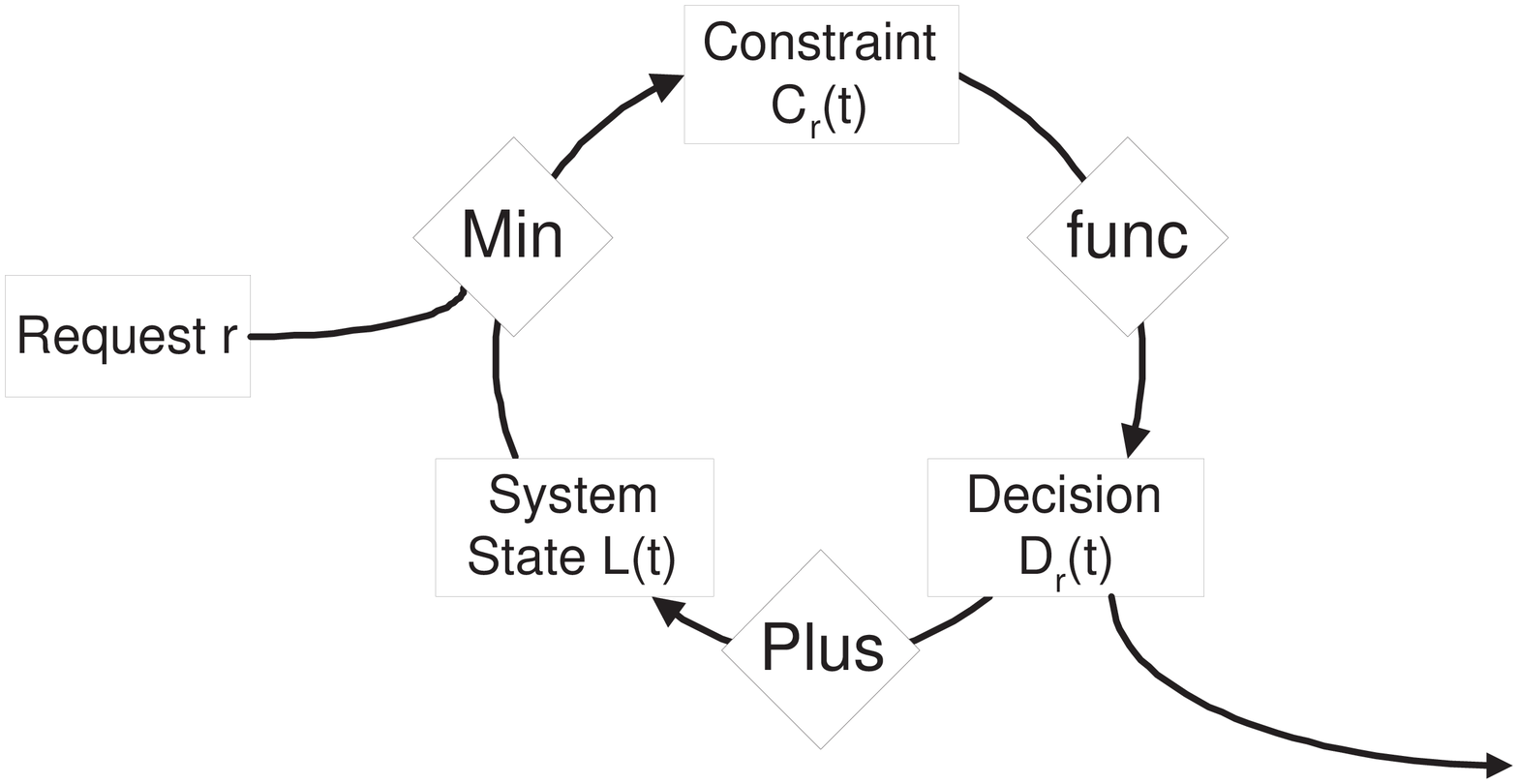}
\caption{Reservation schemes algorithm framework}
\label{logicloop}
\end{figure}

A bandwidth scheduler makes decision for request based on system
state $L(t)$ and request specification $r$. As shown in Figure
\ref{logicloop}, a scheduler first calculates constraint function
$C_r(t)$ for the reservation, considering both request
specification and current system state $L(t)$. Calculation of
constraint is a \emph{min} operation over \emph{time-rate
function} which will be defined below. Constraint function
$C_r(t)$ then is used to make reservation decision $D_r(t)$.
$D_r(t)$ is the output of scheduler, and is also used internally
to update link state $L(t)$.

\subsection{Time-rate function algebra}

We denote the set of all time-rate functions as $\mathcal{F}$, and
we define \emph{Min-Plus} algebra over $\mathcal{F}$:

\begin{equation}(f_1 \;min\; f_2)(t) = min(f_1(t), f_2(t))\end{equation}
\begin{equation}(f_1 + f_2)(t) = f_1(t) + f_2(t)\end{equation}

While Min-algebra is a semigroup, Plus-algebra is a group with
identity element $f^0(t)=0,\forall t \in (-\infty,\infty)$. We
define $\leq$ relation over $\mathcal{F}$ as:

\begin{equation}f_1 \leq f_2\textrm{, iff }f_1(t) \leq f_2(t),\forall t \in (-\infty,\infty)\end{equation}

Note that $\mathcal{F}$ with $\leq$ is a partial order set not
satisfying comparability condition.

$a_r, d_r, R^{max}_r$ in request specification determines a
time-rate function, which can be viewed as the original constraint
function imposed by request specification:

\begin{equation}C_r^{request}(t)=R^{max}_r h(t-a_r) - R^{max}_r h(t-d_r)\end{equation}

where:
\begin{equation}h(t) = \left\{ \begin{array}{ll}
1 & t \in [0, \infty)\\
0 & \textrm{otherwise}\\
\end{array} \right.\end{equation} is the Heaviside step function (unistep function). Translation of $h(t)$ is indicator function for half-open interval.

The constraint calculation stage shown in Figure \ref{logicloop}
is to consider both $C_r^{request}(t)$ and system reservation
state $L(t)$, so that the resulted $C_r(t)$ returns the maximum
bandwidth that can be allocated to request $r$ at time $t$:

\begin{equation}C_r(t) = (C_r^{request} \;min\; L_1 \;min\; L_2 \;min\; \dots \;min\; L_k)(t)\end{equation}

where we assume links $L_1, L_2, \dots L_k$ form path from
$source[r]$ to $dest[r]$, and $L_i(t)$ is the time-(remained
bandwidth) function for link $L_i$. The \emph{min} operation is
illustrated in Figure \ref{fig:constraint} with two links $L_1$
and $L_2$ in request $r$'s path:

\begin{figure}[!ht]
\centering
\includegraphics[height=50mm]{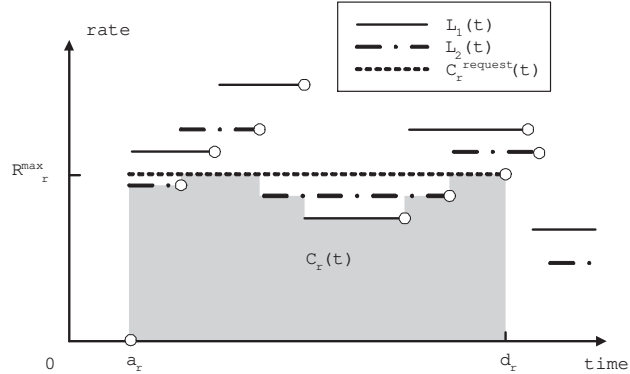}
\caption{Calculate request's constraint function $C_r(t)$}
\label{fig:constraint}
\end{figure}

Reservation decision function $D_r(t)$ returns the reserved data
rate for $r$ at time $t$.  If scheduler rejects the request, no
bandwidth will be reserved for the request in the whole time axis.
Thus rejection decision can be represented by $f^0$. $D_r(t)$
satisfies:

\begin{equation}\label{eq:constraint1} D_r(t) \leq C_r(t) \end{equation}
\begin{equation}\label{eq:constraint2} \int_{t=a_r}^{d_r} D_r(t) dt = v_r\textrm{, if }D_r \neq f^0\end{equation}

In the system state update stage shown in Figure \ref{logicloop}:

\begin{equation}L_i(t) = (L_i - D_r)(t),\forall L_i\in \textrm{path of }r\end{equation}

At time $\tau$, an empty link $L_i$ without any reservation has
$L_i(t) = B[L_i] h(t-\tau)$, where $B[L_i]$ is the total capacity
of link $L_i$.

\subsection{Step time-rate functions}
\label{complexity}

General time-rate functions are not suitable for implementation,
thus we restrict our discussion to a special class of time-rate
functions, i.e., the \emph{step time-rate functions}, which are
easy to be stored and processed.

Formally, a function is called \emph{step function} if it can be
written as a finite linear combination of indicator functions of
half-open intervals. Informally speaking, a step function is a
piecewise constant function having only finitely many pieces. A
time step function $f(t)$ can be represented as:
\begin{equation}f(t) = a_1 h(t - b_1) + a_2 h(t - b_2) + \cdots + a_n h(t - b_n)\end{equation}
We denote the set of all step functions as $\mathcal{F}_s \subset
\mathcal{F}$. A step function with $n$ non-continuous points can
be uniquely represented by a $2\times n$ matrix $\left[
\begin{array}{ccc}
a_1 & \ldots & a_n \\
b_1 & \ldots & b_n \end{array} \right]$ with elements in first row
non-zero, and elements in second row strictly increasing. All step
functions with $n$ non-continuous points form \emph{$n$-step
function set} $\mathcal{F}^n_s \subset \mathcal{F}_s$.
$\mathcal{F}^0_s=\{f^0\}$. $\mathcal{F}^1_s=\{\textrm{all
translations of }h(t)\}$. All non-regressive linear combination of
two different elements in $\mathcal{F}^1_s$ form
$\mathcal{F}^2_s$. For $f^2\in \mathcal{F}^2_s$, if $a_1+a_2=0$,
$f^2$ and $f^0$ encompass a rectangular in time-rate coordinate.
All such $f^2$ form the \emph{rectangular function set}
$\mathcal{F}_{rec}$. We also define \emph{general $n$-step
function set} $\mathcal{G}^n = \mathcal{F}^0 \cup \mathcal{F}^1
\ldots \mathcal{F}^n$.

Following discussions restrict reservation schemes to make
decision in step function form, i.e., $D_r\in \mathcal{F}_s$ . For
$f^n(t) \in \mathcal{F}^n_s$ and $f^m(t) \in \mathcal{F}^m_s$, it
is easy to show that $(f^n \;min\; f^m)(t) \in
\mathcal{F}^{n+m}_s$, and $(f^n + f^m)(t) \in
\mathcal{F}^{n+m}_s$, i.e., both \emph{min} and \emph{plus}
operations are closed in $\mathcal{F}_s$, thus constraint function
$C_r(t)$ and time-(remained bandwidth) function $L_i(t)$ are also
step functions. The computation and space complexity for
\emph{min}, \emph{plus} and \emph{order} operations over function
$f^n(t)$ and $f^m(t)$ are $O(n+m)$. We discuss calculation of
$D_r(t)$ based on $C_r(t)$ and $v_r$ in next section.
\newpage

\section{Reservation schemes}
\label{sec:schemes}
\subsection{Schemes taxonomy and heuristics}
Existing RSVP-type reservation schemes only supports reservation
of a fixed bandwidth from a fixed start time, which we name as
\emph{FixTime-FixRate} schemes. Slightly more general are
\emph{FixTime-FlexRate} schemes, which still enforces a fixed
start time, but allow scheduler to flexibly determine the
reservation bandwidth. To further generalize the idea, we have
\emph{FlexTime-FlexRate} schemes, which allows reservation starts
from any time in $[a_r, d_r]$ and reserves any rate (but need to
be constant) continuously until transfer completes. Finally, by
allowing reservation comprise of multiple ($n\leq1$) sub-intervals
with different reservation bandwidths, we have
\emph{Multi-Interval} schemes. Regarding their solution space,
\emph{FixTime-FixRate} $\subset$ \emph{FixTime-FlexRate} $\subset$
\emph{FlexTime-FlexRate} $\subset$ \emph{MultiRate}. Their
different flexibilities are summarized in Table
\ref{table:schemes}.

\begin{table}
\begin{center}
{\small
\begin{tabular}{|c|p{3.5cm}|c|}
\hline
Schemes & accept decision & flexibility \\
\hline
FixTime-FixRate & $D_r(t)=C_r(t)$ & 0\\
\hline
FixTime-FlexRate & $D_r(t)\in \mathcal{F}_{rec}$ with term $h(t-a_r)$ & 1\\
\hline
FlexTime-FlexRate & $D_r(t)\in \mathcal{F}_{rec}$ & 2\\
\hline
Multi-Interval & $D_r(t)\in \mathcal{G}_{n}$ & 2n-2 \\
\hline
\end{tabular}
} \caption{Reservation schemes} \label{table:schemes}
\end{center}
\end{table}

The flexibility makes it hard to choose a suitable decision
$D_r(t)$ if multiple candidates are available. As mentioned in
Section \ref{sec:motivation}, there are multiple performance
criteria, increasing accept probability, minimizing flow time, and
ensuring fairness among flows, just name a few. In fact, even for
RSVP-type reservation scheme with only two choices (reject, or
accept the request with fixed rate at fixed start time), it is
hard to make an optimal selection as proved in
\cite{marchal05sharing}. Instead, we use simple heuristics to
select representative scheme from each family for performance
comparison. A \emph{threshold-based rate-tuning} heuristic is used
to choose candidate from \emph{FixTime-FlexRate} schemes which
will be detailed in Section \ref{sec:eval}. Simple {Greedy-Accept}
and {Minimize-FlowTime} heuristics are used to choose candidate
from \emph{FlexTime-FlexRate} family and \emph{Multi-Interval}
family.

\emph{Greedy-Accept} means: If there is at least one feasible
solution to accept a coming request, the request should not be
rejected. Greedily accept new request is not optimal in an
off-line sense, because sometimes it maybe better to
\emph{Early-Reject} a request even when feasible solution exists,
so that capacity can be kept for more rewarded-requests which
arrive later. Despite this, it is an interesting heuristic to
study, because:
\begin{itemize}
\item \emph{Greedy-Accept} heuristic can be used orthogonally with
trunk reservation to mimic the behavior of \emph{Early-Reject};
\item \emph{Greedy-Accept} introduces a strict priority based on
arriving order, which by itself is a reasonable assignment
philosophy.
\end{itemize}

\emph{Minimize-FlowTime} means: If there are multiple feasible
solutions in the solution space, the one with minimal completion
time will be chosen. Besides the straightforward benefit on
minimizing flow time, this philosophy also helps maximize the
utilization of resource in near future, which otherwise is more
likely to be wasted if no new request comes soon. However, since
the near future is more densely packed with reservation, assuming
all requests have identical $R_{exp}$, then a small volume request
with short life span is easier to get rejected than a large volume
request with long life span. This unfairness can also be addressed
by volume-based trunk reservation.

\subsection{\emph{FixTime-FixRate} schemes}

In \emph{FixTime-FixRate} schemes, request specifies its desired
reservation rate. Scheduler can only decide to accept or reject.
As shown in \cite{altman02}, reducing reservation rate increases
system's Erlang capacity. Thus a candidate \emph{FixTime-FixRate}
scheme to maximize accept rate is to enforce:
\begin{equation}
D_r=\left\{ \begin{array}{ll}
R_r^{min} (h(t-a_r) - h(t- d_r)) & \textrm{if } R_r^{min} \leq C_r(a_r) \\
f^0 & \textrm{otherwise}\\
\end{array} \right.
\end{equation}

Here $R_r^{min} = \frac{v_r}{d_r-a_r}$ satisfy Equation
(\ref{eq:constraint2}). In this scheme, every accepted request
completes transfer exactly at its deadline, if a dull transfer
protocol is used. This is the reservation scheme used in Figure
\ref{img:multiplexing}. Notice that for \emph{FixTime} schemes
without advance reservation, Equation (\ref{eq:constraint1}) is
simplified to consider constraint function $C_r(t)$'s value at
$a_r$ only, because:
\begin{itemize}
\item \emph{FixTime} schemes' reservation is enforced to begin
from $a_r$; \item Under \emph{FixTime} schemes without advance
researvation, time-(remained capacity) function $L_i(t)$ for any
link $L_i$ is non-decreasing along time axis.
\end{itemize}

\subsection{\emph{FixTime-FlexRate} schemes}
\label{now}

\emph{FixTime-FlexRate} schemes still enforce transfer start at
$a_r$, thus $D_r(t)\in \mathcal{F}_{rec}$ must have term
$h(t-a_r)$. Compared to \emph{FixTime-FixRate} schemes,
\emph{FixTime-FlexRate} schemes can flexibly choose the rate
parameter $R_r$ in $D_r(t)$. \emph{FixTime-FlexRate} schemes
allocate a single rate $R_r$ for accepted request $r$ from its
arrival time $a_r$ to its completion time $a_r + \frac{v_r}{R_r}$:
\begin{equation}
D_r(t) = R_r (h(t-a_r) - h(t - a_r - \frac{v_r}{R_r}))
\end{equation}

The second term above is calculated using Equation
(\ref{eq:constraint2}). While Equation (\ref{eq:constraint1}) is
simplified as: $a_r + \frac{v_r}{R_r} \leq d_r$ thus $R_r \geq
\frac{v_r}{d_r-a_r}$, and $R_r \leq C_r(a_r)$ similar to
\emph{FixTime-FixRate} schemes.

\subsection{\emph{FlexTime-FlexRate} schemes}
\emph{FlexTime-FlexRate} schemes relax the fix start time
constraint. Thus, \emph{Decision Function} $D_r(t)$ of
\emph{FlexTime-FlexRate} schemes can be any rectangular function
satisfying Equation (\ref{eq:constraint1}) and
(\ref{eq:constraint2}). \emph{FlexTime-FlexRate} schemes allocate
a single rate $R_r$ in interval $[t_r^{start}, t_r^{start} +
\frac{v_r}{R_r}] \subseteq [a_r, d_r]$. The $D_r$ can be fully
characterizes by a pair $(t_r^{start}, R_r)$. Completion time is
calculated using Equation (\ref{eq:constraint2}).

To simplify Equation (\ref{eq:constraint1}), we define
\emph{constraint rectangular function set}
$\mathcal{F}_{rec}^{constraint}$ and \emph{Pareto optimal
rectangular function set} $\mathcal{F}_{rec}^{Pareto}$ for
constraint function $C_r(t)$:

\begin{equation}\mathcal{F}_{rec}^{constraint}=\{f(t) | f(t) \in {F}_{rec} \textrm{ and } f(t) \leq C_r(t)\}\end{equation}
\begin{eqnarray}\mathcal{F}_{rec}^{Pareto} & = & \{f(t) | f(t) \in {F}_{rec}^{constraint} \textrm{and}\nonumber \\
& & !\exists\; g(t) \in \mathcal{F}_{rec}^{constraint}, g(t) >
f(t) \}\end{eqnarray}

\begin{figure}[!ht]
\centering
\includegraphics[height=50mm]{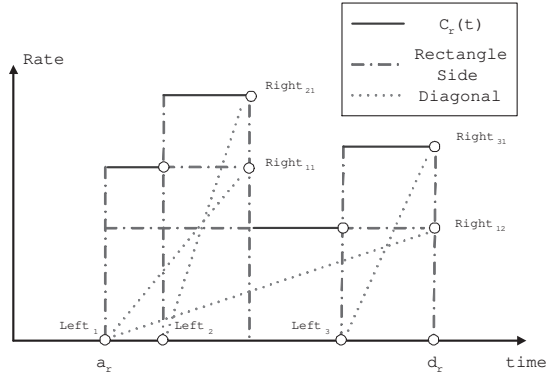}
\caption{Pareto Optimal Rectangular function set}
\label{img:paretoOptRect}
\end{figure}

\emph{Pareto optimal rectangular function set} of a n-step
constraint function $C_r(t)$ can be calculated in $O(n^2)$ as
illustrated in Figure \ref{img:paretoOptRect},
$\mathcal{F}_{rec}^{Pareto}$ contains $O(n^2)$ elements.

Apply \emph{Greedy-Accept} and \emph{Minimize-FlowTime} heuristics
here: a request $r$ is rejected, if and only if there is no $f(t)
\in \mathcal{F}_{rec}^{Pareto}$ with integration no less than
$v_r$; otherwise, all Pareto optimal rectanglar functions with
large enough integration are checked to identify the one providing
minimum flow time. Given a Pareto optimal rectangular function
$f(t) = a_1 (h-t_1) - a_1 (h-t_2)$, the minimum flow time it can
provide is $t_1 + \frac{v_r}{a_1}$. The implementation of this
scheme is detailed in Table \ref{SR-MaxPack-MinDelay}.

\begin{table}
\begin{center}
{\small
\begin{tabular}{|p{13cm}|}
\hline
struct time-rate\{\\
$\qquad$double time;\\
$\qquad$double rate;\\
$\qquad$boolean unVisited = true;\\
\};\\
\\
Input: 6-tuple representation of request $r$ and its constraint function $C_r(t)$, which is a $n$-step function represented by a time-rate vector $v$. For $i \!\in\! [0,\dots, n-1]$:\\
$\qquad$v[i].time is the $(i+1)^{th}$ noncontinuous points of $C_r(t)$,\\
$\qquad$v[i].rate = $C_r(v[i]$.time$)$.\\
\\
Output: decision d in a time-rate structure. \\
\\
int nextIncrease(int i)\{\\
$\qquad$for(i++; i $<=$ n; i++)\\
$\qquad \qquad$if(v[i-1] $<$ v[i])\\
$\qquad \qquad \qquad$break;\\
$\qquad$return i;\\
\}\\
\\
int nextDecrease(int i)\{\\
$\qquad$if(v[i].unVisited)\{\\
$\qquad \qquad$v[i].unVisited = false;\\
$\qquad \qquad$double r = v[i].rate;\\
$\qquad \qquad$for(i++; i $<$ n; i++)\\
$\qquad \qquad \qquad$if(r $>$ v[i].rate)\\
$\qquad \qquad \qquad \qquad$break;\\
$\qquad \qquad \qquad$return i;\\
$\qquad$\}\\
$\qquad$else\\
$\qquad \qquad$return n;\\
\}\\
\\
struct time-rate reservation(request r, struct time-rate v[])\{\\
$\qquad$struct time-rate d;\\
$\qquad$d.time = r.deadline;\\
$\qquad$d.rate = 0;\\
$\qquad$for(int left = 0; left $<$ n-1 \&\& v[left].rate $>$ 0 \&\& v[left].time $<$ d.time; left = nextIncrease(left))\{\\
$\qquad \qquad$double resv-rate = v[left].rate;\\
$\qquad \qquad$for(int right = nextDecrease(left); right $<$ n; right = nextDecrease(right))\{\\
$\qquad \qquad \qquad$if(v[left].time + r.volume / resv-rate $<$ d.time)\{\\
$\qquad \qquad \qquad \qquad$d.time = v[left].time + r.volume / resv-rate;\\
$\qquad \qquad \qquad \qquad$d.rate = resv-rate;\\
$\qquad \qquad \qquad \qquad$break;\\
$\qquad \qquad \qquad$\}\\
$\qquad \qquad \qquad$resv-rate = v[right].rate;\\
$\qquad \qquad$\}\\
$\qquad$\}\\
$\qquad$if(d.rate $>$ 0) d.time $-=$ r.volume / d.rate;\\
$\qquad$return d;\\
\}\\
\hline
\end{tabular}
} \caption{\emph{Greedy-Accept} \emph{Minimize-FlowTime}
\emph{FlexTime-FlexRate} schemes} \label{SR-MaxPack-MinDelay}
\end{center}
\end{table}

\subsection{\emph{Multi-Interval} schemes}

Compared to all above schemes, reservation decision in
\emph{Multi-Interval} schemes can be composed of multiple
intervals with different reservation rates. Note that
\emph{Multi-Interval} schemes are different from preemptive
schemes. Although multiple rates can be used in
\emph{Multi-Interval} schemes, and flows are probably scheduled to
transfer in two discontinuous intervals, this decision is
determined at the moment the request arrives, and is not changed
(preempted) after that.

\begin{figure}[!ht]
\centering
\includegraphics[height=40mm]{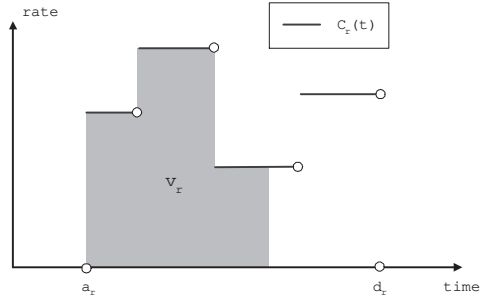}
\caption{\emph{Multi-Interval} schemes} \label{mr_fig}
\end{figure}

Apply \emph{Greedy-Accept} and \emph{Minimize-FlowTime} heuristics
here, if integration of $C_r(t)$ over time axis is larger than
$v_r$:
\begin{equation}D_r(t)= \left\{ \begin{array}{ll}C_r(t) & t \leq \tau\\
0 & t > \tau\\
\end{array} \right. \end{equation}
where time $\tau$ satisfies: $\int_{t=a_r}^{\tau} C_r(t) dt =
v_r$. $D_r(t)=f^0$ if no such $\tau$ exists. As shown in Figure
\ref{mr_fig}, when $C_r(t) \in \mathcal{F}^n_s $ is a $n$-step
function, computational complexity of MR-MaxPack-MinDelay scheme
is $O(n)$, and $D_r(t) \in \mathcal{G}^n_s$.

Sometimes it is useful to enforce $D_r(t) \in \mathcal{G}^n_s$ for
a constant $n$. For example, \emph{FlexTime-FlexRate} schemes are
subset of \emph{Multi-Interval} schemes enforcing $D_r(t) \in
\mathcal{G}^2_s$. If reservation decision is allowed to be
composed of at most two adjecent subintervals with different
rates, it can be modeled as subset of \emph{Multi-Interval}
schemes enforcing $D_r(t) \in \mathcal{G}^3_s$.
\newpage

\section{Performance evaluation}
\label{sec:eval}

\subsection{Simulation setup}

We use simulation to demonstrate the potential performance gain
from the increasing flexibility. We consider the performance of
both \emph{blocking probability} and \emph{mean flow time} for
following schemes:
\begin{itemize}
\item \emph{FixTime-FixRate-$R_{max}$} scheme is a
\emph{FixTime-FixRate} scheme with reservation rate of $R_{max}$;
\item \emph{FixTime-FixRate-$R_{min}$} scheme is a
\emph{FixTime-FixRate} scheme with reservation rate of $R_{min}$;
\item \emph{Threshold-FixTime-FlexRate} scheme is a simple
\emph{FixTime-FlexRate} scheme which reserves $R_{max}$ when the
minimum unreserved bandwidth among all links along the path is
above a threshold (set as $20\%$ of link capacity in this
simulation), and reservates $R_{min}$ otherwise; \item
\emph{Greedy-Accept} and \emph{Minimize-FlowTime} heuristic in the
\emph{FlexTime-FlexRate} family; \item \emph{Greedy-Accept} and
\emph{Minimize-FlowTime} heuristic in the \emph{Multi-Interval}
family.
\end{itemize}
For all above settings, \emph{dull transport protocol} is assumed,
which uses and only uses reserved bandwidth.

To simplify the discussion on the potential gain of increasing
flexibility, we ideally assume that bulk data transfer requests
arrive online according to a Poisson process with parameter
$\lambda$, all requests have the same volume $v$, $R_{max}=C/10$
and $R_{min}=C/20$, where $C$ is the link capacity. Observation in
this simple setting also helps explain the system behavior in more
general settings, which may have different arrival process, volume
distribution, $R_{max}$ and $R_{min}$.

\subsection{Single Link setting}

We first consider the case of single bottleneck link. Performance
of above schemes is plotted under increasing load.

\begin{figure}[!ht]
\centering
\includegraphics[height=60mm]{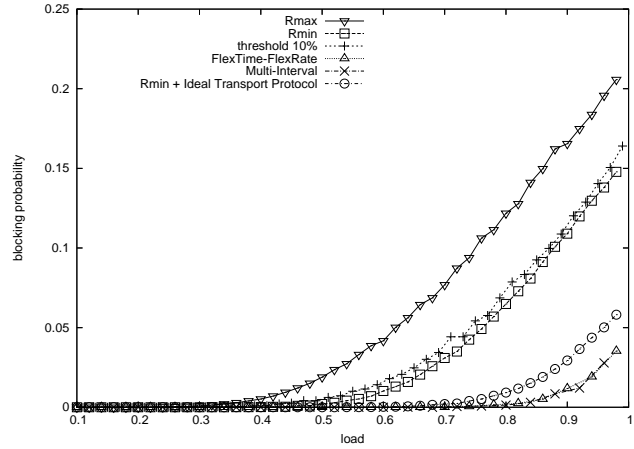}
\caption{Blocking probability of reservation schemes}
\label{img:blocking1}
\end{figure}

\begin{figure}[!ht]
\centering
\includegraphics[height=60mm]{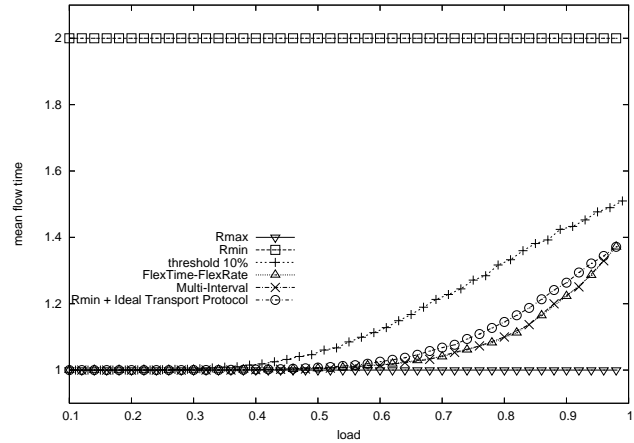}
\caption{Mean flow time of reservation schemes}
\label{img:flowtime1}
\end{figure}

Figure \ref{img:blocking1} shows that in terms of blocking
probability, \emph{FixTime-FixRate-$R_{min}$} scheme performs
better than \emph{FixTime-FixRate-$R_{max}$} scheme. When
reservation rate decreases, two conflicting effects happen: On one
hand, more requests can be accepted simultaneously; on the other
hand, each request takes a longer time to finish. \cite{altman02}
shows that decreasing reservation rates increase system's Erlang
capacity, which is verified in this Figure. However, as
\emph{FixTime-FixRate-$R_{min}$} always conservatively reserve
$R_{min}$, its request flow time is always $v_r / R_{min}$.
Contrarily, flow time of \emph{FixTime-FixRate-$R_{min}$} scheme
is aways $v_r / R_{max}$, which is only half of $v_r / R_{min}$
under our simulation setting, as shown in Figure
\ref{img:flowtime1}.

Exploiting the flexibility of selecting reservation rates,
\emph{Threshold-FixTime-FlexRate} scheme strikes a good balance
between reducing blocking probability and minimizing mean flow
time. When load is low, a new request reserves full rate
$R_{max}$, so that its flow time is minimized. Although the new
request agressively seizes bandwidth, the threshold statistically
ensures that there are still abundant bandwidth left. Thus the
probability is low that in a near future coming flows are blocked
due to this aggressive request. Instead, the new request exploits
the resource which will otherwise be wasted, and also it is able
to release network resource more quickly, which benefits the
system at a middle-range time scale. In the lightly-loaded region
\emph{Threshold-FixTime-FlexRate} scheme performs similar to
\emph{FixTime-FixRate-$R_{max}$} scheme. However when load
increases, links are often run in saturated state, a new request
has higher probability to find remained capacity below threshold.
Thus in this region, \emph{Threshold-FixTime-FlexRate} scheme
automatically adapts its behavior to perform similar to
\emph{FixTime-FixRate-$R_{min}$}. From the two figures, it is
observed that \emph{Threshold-FixTime-FlexRate} scheme has a much
lower blocking probability than \emph{FixTime-FixRate-$R_{max}$}
scheme, while has a much lower mean flow time than
\emph{FixTime-FixRate-$R_{min}$} scheme.

In this single link setting, behavior of selected
\emph{FlexTime-FlexRate} and \emph{Multi-Interval} schemes are
identical. This is an artificial result of the uniform volume and
$R_{max}$ setting, as well as the integer value of $C/R_{max}$. We
also conduct extensive simulations over more general volume,
$R_{max}$ and $R_{min}$ distribution over a single link, and
results also show that the performance of \emph{FlexTime-FlexRate}
and \emph{Multi-Interval} remains close. Both
\emph{FlexTime-FlexRate} and \emph{Multi-Interval} schemes perform
much better than above three schemes in both blocking rate and
flow time.

A remarkable observation is that, \emph{FlexTime-FlexRate} and
\emph{Multi-Interval} schemes with dull transport protocol even
outperform the \emph{FixTime-FixRate-$R_{min}$} scheme equipped
with ideal transport protocol, in terms of both blocking rate and
flow time (see the \emph{Rmin + Ideal Transport Protocol} curve in
both Figure. In addition, the small flow time of \emph{Rmin +
Ideal Transport Protocol} is achieved opportunistically by ideal
transport protocol, which can not be guaranteed at the moment when
the reservation is made (in contrast,
\emph{FixTime-FixRate-$R_{min}$} scheme can only guarantee that
accepted requests are completed before deadline). Thus other
co-allocated resources can not exploit the small flow time to
increase their scheduling efficiency. On the other hand, the
request flow time is known and guaranteed in reservation schemes
at the moment when request is processed. This predictability can
benefit other co-allocated resources. This result strongly
motivates the study of advanced reservation schemes.
\newpage

\subsection{Grid network setting}

\begin{figure}[!ht]
\centering
\includegraphics[height=50mm]{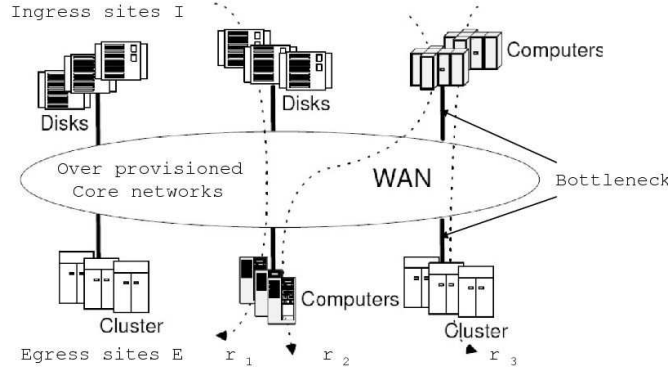}
\caption{Topology} \label{topology}
\end{figure}

We also evaluate different schemes' performance in a network
setting. We use the topology as shown in Figure \ref{topology}.
$n$ ingress sites and $n$ egress sites are interconnected by
over-provisioned core networks. Each site composed of a cluster of
grid nodes, and is connected to core network with a link of
capacity $C$. The maximal aggregate bandwidth demands from the
culster may exceed $C$, making these links potential bottlenecks.
For simplicity, we assume that the core network is
over-provisioned, like the visioned Grid5000 networks in France
\cite{grid5000}. Core network can be provisioned, for example,
using hose model \cite{duffield99}. When generating request, its
source is randomly selected from ingress sites, then a random
destination is selected independently among egress sites. All
sites have the same probability to be chosen.

\begin{figure}[!ht]
\centering
\includegraphics[height=60mm]{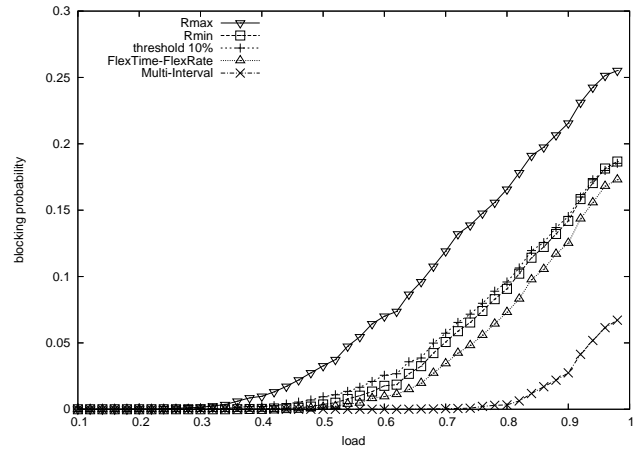}
\caption{Blocking probability of reservation schemes}
\label{img:blocking10}
\end{figure}

\begin{figure}[!ht]
\centering
\includegraphics[height=60mm]{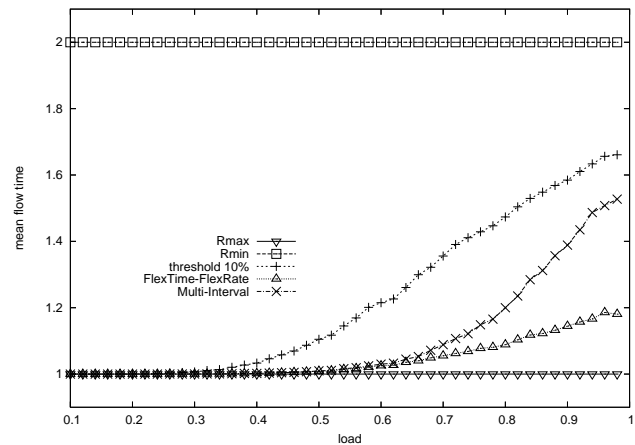}
\caption{Mean flow time of reservation schemes}
\label{img:flowtime10}
\end{figure}

Figure \ref{img:blocking10} and Figure \ref{img:flowtime10} plot
the performance when there are $10$ ingress nodes and $10$ egress
nodes in the network. Compared to Figure \ref{img:blocking1} and
Figure \ref{img:flowtime1}, three phenomenons are observed:

\begin{itemize}
\item Overall, performance of schemes degrades slightly; \item
\emph{FlexTime-FlexRate} scheme's blocking probability shows a big
increase, and its performance is no longer close to
{Multi-Interval} scheme; \item \emph{Multi-Interval} scheme's mean
flow time performance deteriorates obviously.
\end{itemize}

The overall performance degration can be traced to the fact that
reservation in a network need to consider multiple links (both
ingress and egress link in this topology). A reservation request
is blocked or its flow time becomes longer when any one of them is
congested. If we assume that congestion states in two links are
independently and identically distributed, with mean congestion
probability $p$, the probability that there is at least one of
them being congested is $2p-p^2 > p$. This intuitively explains
the overall degration of performance.

\begin{figure}[!ht]
\centering
\includegraphics[height=40mm]{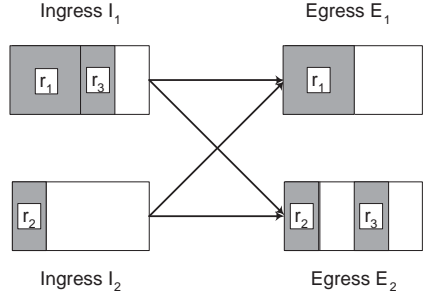}
\caption{A fragmentation example} \label{img:fragmentation}
\end{figure}

The performance degration of \emph{FlexTime-FlexRate} scheme's
blocking probability and \emph{Multi-Interval} scheme's mean flow
time can be explained using a simple example in Figure
\ref{img:fragmentation}.

In this example, there are two ingress links and two egress links
interconnected by over-provisioned core networks. Existing request
$r_1$ reserves bandwidth in $I_1$ and $E_1$, while existing
request $r_2$ reserves bandwidth in $I_2$ and $E_2$ as shown in
the Figure. At current system time, a new request $r_3$ arrives at
$I_1$ with destination $E_2$. For the three \emph{FixTime} schemes
(\emph{FixTime-FixRate-$R_{max}$} scheme,
\emph{FixTime-FixRate-$R_{min}$} scheme and
\emph{Threshold-FixTime-FlexRate} scheme), they are not allowed to
accept $r_3$ since bandwidth is fully reserved for the current
time. This prevents fragmentation as shown in the Figure when both
\emph{FlexTime-FlexRate} scheme and \emph{Multi-Interval} scheme
exploit their flexibility to accept $r_3$. This time-axis
framentation increases \emph{FlexTime-FlexRate} scheme's blocking
probability, since \emph{FlexTime-FlexRate} scheme can only
allocate a continuous time interval. On the other hand, blocking
rate of \emph{Multi-Interval} scheme is not affected as much as
\emph{FlexTime-FlexRate} scheme because \emph{Multi-Interval}
scheme can make use of multiple (discontinuous) intervals. However
\emph{Multi-Interval} scheme's mean flow time is affected.

\begin{figure}[!ht]
\centering
\includegraphics[height=60mm]{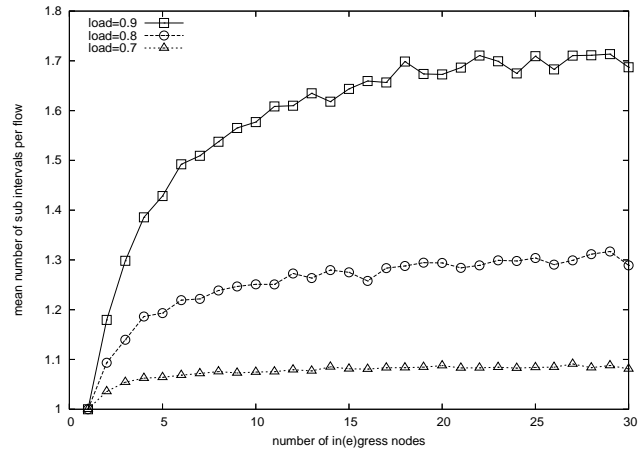}
\caption{Mean number of intervals per flow in
\emph{Multi-Interval} scheme} \label{img:interval_count}
\end{figure}

In above examples, \emph{Multi-Interval} schemes often give the
best perfromance. However, using multiple intervals comes at a
cost. Figure \ref{img:interval_count} shows the increase trend of
sub-interval number when network size is increased. It is shown
that this number becomes quite stable around a small level, when
the number of nodes grows larger than the multiplexing level of a
single link, which is $C/R_{max}$. This result holds for different
load levels. This observation shows the feasibility of exploiting
\emph{Multi-Interval} scheme.

\newpage

\section{System architecture}
\label{sec:architecture}

The logic framework shown in Figure \ref{logicloop} corresponds to
a centralized scheduler, which may not be desirable because:
\begin{itemize}
\item links may be under control of different authorities; \item
when network size grows, the centralized scheduler itself may
become a bottleneck; \item Centralized scheduler presents an
one-failure-point.
\end{itemize}

\begin{figure}[!ht]
\centering
\includegraphics[height=50mm]{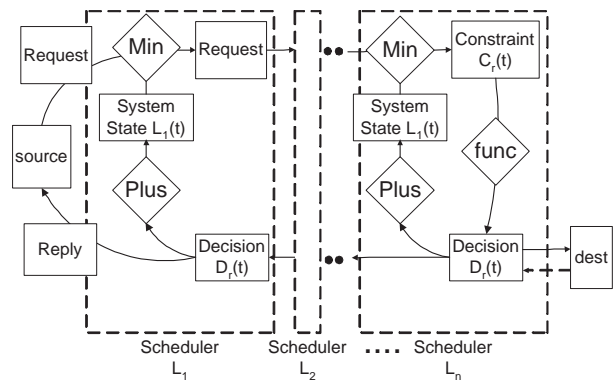}
\caption{Distributed architecture} \label{distributed}
\end{figure}

Thus we present a simple distributed architecutre as shown in
Figure \ref{distributed}. In this architecture, every bottleneck
link is associated with a local bandwidth scheduler, which
maintains the local reservation state. Request generated from the
source first arrives at link $L_1$, whose scheduler uses $min$
operation to combine its local link state constraint into the
request specification. The updated request specification is
forwards to the nexthop. In this way, the constraint function is
updated hop by hop: $C_r^{i}(t) = (L_i \;min\; C_r^{i-1})(t)$.
When request reaches the last hop $L_n$, the constraint function
$C_r(t)$ is completely constructed, and the scheduler in $L_n$
makes decision $D_r(t)$ based on $C_r(t)$. $D_r(t)$ is sent to
destination, which may issue a confirmation. $D_r(t)$ is then sent
through the same path back to source. $D_r(t)$ is kept unchanged
along the path, and each hop uses $D_r(t)$ to update its local
reservation state $L_n$.

Single out a local scheduler, its logic can still be interpreted
using the logic framework of Figure \ref{logicloop}. The only
difference is that for schedulers not in the last hop, their
\emph{``func''} operation is not a local operation but depends
recursively on the next hop.
\newpage

\section{Related works}
\label{sec:relate}

Admission control and bandwidth reservation have been studied
extensively in multimedia networking. A real-time flow normally
requests a specified value of bandwidth. Existing reservation
schemes such as RSVP \cite{braden97} attempt to reserve the
specified bandwidth immediately when request arrives. Reservation
remains in effect for an indefinite duration until explicit
``Teardown'' signal is issued or soft state expires. No
time-indexed reservation state is kept.

Time-indexed reservation is needed when considering advance
reservation of bandwidth \cite{wischik98}, which allows requesting
bandwidth before actual transfer is ready to happen. For example,
a scheduled tele-conference may reserve bandwidth for a specified
future time interval. \cite{burchard03} shows that advance
reservation causes bandwidth fragmentation in time axis, which may
significantly reduce accept probability of requests arriving
later. To address the problem, they propose the concept of
\emph{malleable reservation}, which defines advance reservation
request with flexible start time and rate.

Optimal control and their complexity is studied for different
levels of flexibility. \cite{altman01} studies call admission
control in a resource-sharing system, i.e. how to use the reject
flexibility regarding different classes of traffic. Optimal policy
structure is identified for some special case.
\cite{marchal05sharing} proved that in a network with multiple
ingress and egress sites, \emph{off-line} optimization of accept
rate for uniform-volume uniform-rate requests with randomly
specified life span is NP-complete. They also consider flexible
tuning of reservation rate. \cite{altman02} studies the increase
of Erlang capacity of a system by decreasing the service rate. In
its essential, such service rate scaling is identical to the
capacity scaling, which is studied by \cite{kelly91} and
\cite{hunt94} to approximate large loss networks.

There is also a large literature of online job scheduling with
deadline, for example, \cite{goldwasser99}, \cite{li05},
\cite{simons83}. A job monopolizes processor for the time it's
being scheduled, which maps exactly to packet level scheduling,
while in flow level, we must consider multiple flows share
bandwidth concurrently, as represented by $R_{max}$.
\newpage
\section{Conclusion}
\label{sec:conclusion} In this paper, we study the bandwidth
reservation problem for bulk data transfers in grid networks. We
model grid networks as multiple sites interconnected by wide area
networks with potential bottlenecks. Data transfer requests arrive
online with specified \emph{volumes} and \emph{deadlines}, which
allow more flexibility in reservation schemes design. We formalize
a general non-preemptive reservation framework, and use simulation
to examine the impact of feasibility over performance. We also
propose a simple distributed architecture for the given framework.
The increased flexibility can potentially improve system
performance, but the enlarged design flexibility also raises new
challenges to identify appropriate reservation schemes inside the
solution space.
\newpage
\tableofcontents

\bibliographystyle{plain}

\end{document}